\documentstyle[floats,graphicx,aps,epsfig]{revtex}
\begin{document}
\draft
%
%
\title{C-axis Penetration Depth and Inter-layer Conductivity in the
Thallium Based Cuprate Superconductors }
\author{Diana Duli\'c, D. van der Marel, A. A. Tsvetkov$^{\ast}$}
\address{Laboratory of Solid State Physics, Materials Science Centre,
Nijenborgh 4, 9747 AG Groningen, The Netherlands}
\author{W. N. Hardy}
\address{Department of Physics and Astronomy, University of British
Columbia, Vancouver, BC, V6T 1Z1, Canada}
\author{Z. F. Ren and J. H. Wang}
\address{Department of Chemistry, SUNY at Buffalo, Buffalo NY
14260-3000, USA}
\author{B. A. Willemsen}
\address{Superconductor Technologies Inc., Santa Barbara,CA 93111-2310}
\date{\today}
\maketitle
\begin{abstract}
The $c$-axis Josephson plasmons in optimally doped single-layer
and bi-layer high T$_c$ cuprates Tl$_2$Ba$_2$CuO$_6$ and
Tl$_2$Ba$_2$CaCu$_2$O$_8$ have been investigated using infrared
spectroscopy. We observed the plasma frequencies for these two
compounds at 27.8 and 25.6 cm$^{-1}$ respectively, which we
interpret as a Josephson resonances across the TlO blocking
layers.
No maximum in the temperature dependence of the $c$-axis
conductivity was observed below T$_c$, indicating that even in the
superconducting state a coherent quasi-particle contribution to
the $c$-axis conductivity is absent or very weak, in contrast to
the behaviour of the $ab$-plane conductivity.
\end{abstract}
\begin{figure}
 \centerline{\epsfig{figure=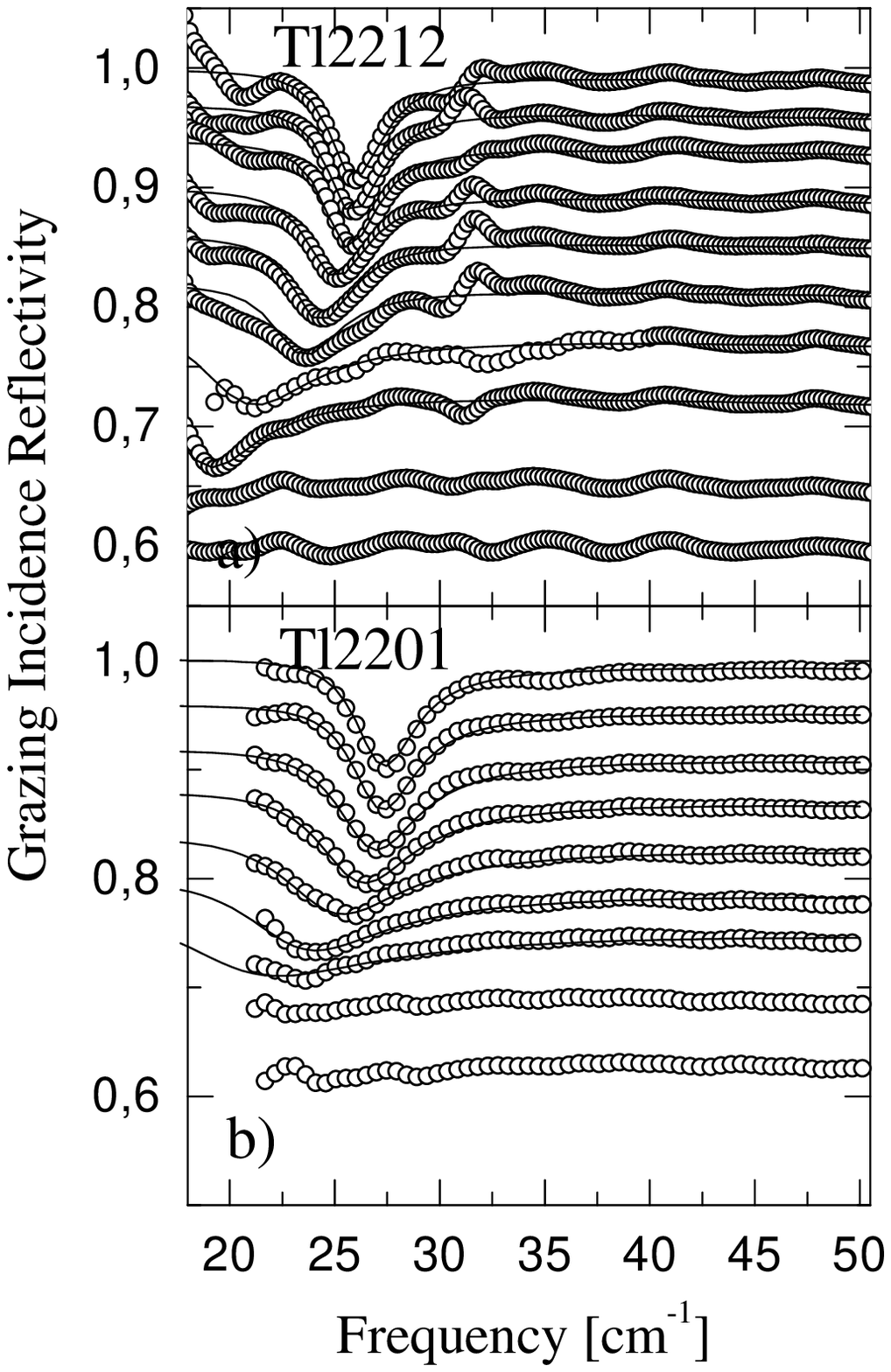,width=9cm,clip=}}
 \caption{
 a) Grazing incidence reflectivity (open circles) of
 Tl2212. From top to bottom: 4K, 10K, 20K, 30K, 40K, 50K, 60K, 70K,
  80K and 110K. The curves have been offset for clarity.
 The solid curves are fits to the data.
 b) Grazing incidence reflectivity (open circles) of Tl2201, at 4K,
 10K, 20K, 30K, 40K, 50K, 60K, 75K, and 90K. The solid curves
 are fits to the data.}
 \label{graz}
\end{figure}
\begin{figure}
 \centerline{\epsfig{figure=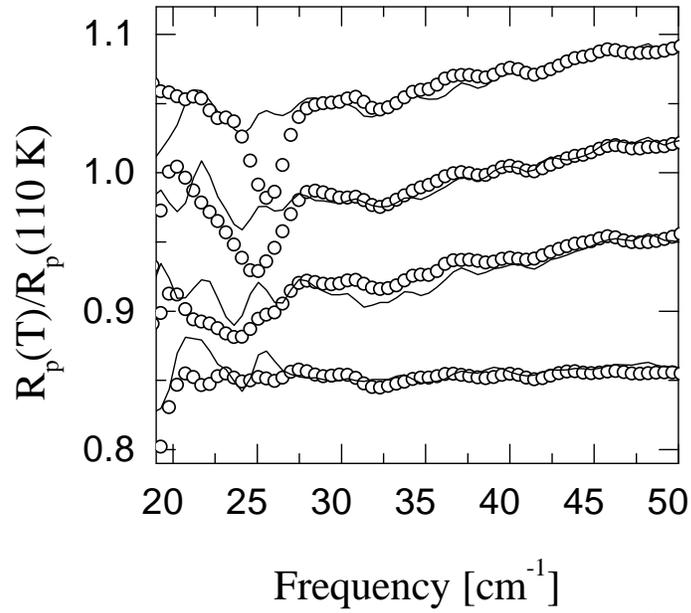,width=9cm,clip=}}
 \caption{
 P-polarized reflectivity normalized to the spectrum at 110 with magnetic
 field (solid curve) and without magnetic field (open circles).
 From top to bottom: 4K, 30K, 60K, 90K.}
 \label{magn}
\end{figure}
\begin{figure}
 \centerline{\epsfig{figure=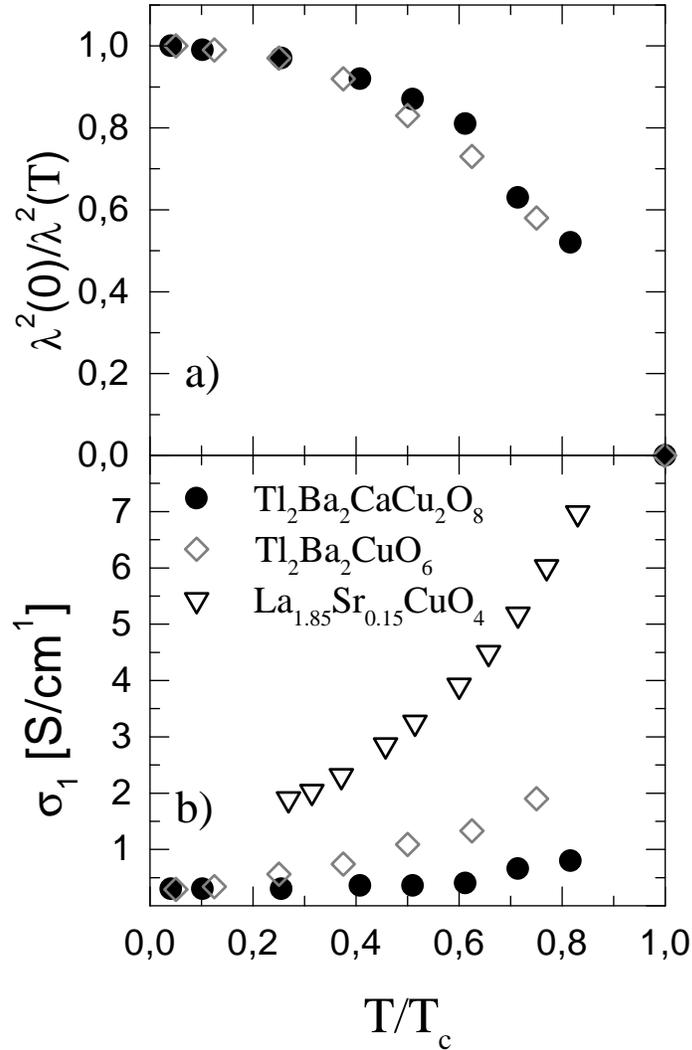,width=9cm,clip=}}
 \caption{
 a) Temperature dependance of the $c$-axis penetration depth of Tl2201
 (open diamonds), and Tl2212 (solid circles).
 b) Temperature dependance of the $c$-axis optical conductivity of
 Tl2201, Tl2212, and LSCO}
 \label{sig}
\end{figure}

Studies of the $c$-axis properties in High-T$_{\text{c}}$
superconductors are of considerable importance. Although the
materials are highly anisotropic, with much of the physics being
of a two dimensional nature, it is clear that an understanding of
both the $c$-axis and $ab$-plane behavior is relevant to the quest
for the mechanism of high-T$_{\text{c}}$ superconductivity.
\\
It is already well established that the normal state c-axis
transport in these materials is strongly incoherent. A number of
theories treat this problem, and generally they can be divided
into two categories, {\em i.e.} Fermi-liquid and non-Fermi-liquid
approaches. The Fermi-liquid approaches rely on special properties
of the quasi-particle scattering, {\em e.g.} a strong momentum
dependence along the quasi two-dimensional Fermi surface of the
quasiparticles \cite{millis}. Within the non-Fermi-liquid
approaches the transport processes involve particles carrying
unconventional spin and charge quantum numbers. One of the most
radical approaches, based on spin-charge separation in the copper
oxide planes, has resulted in the notion of 'confinement' of
single charge carriers to the copper-oxide planes in the normal
state \cite{Anderson}. Within the latter class of models the
formation of Cooper-pairs is accompanied by deconfinement of those
pairs, resulting in a center-of-mass kinetic energy gain.
In principle this also provides a mechanism for superconductivity
\cite{chakravarty}, and a fundamental experimental test of this
hypothesis was proposed by Anderson \cite{ILTtest}. The
experimental results indicated that interlayer tunneling of pairs
is {\em not} the main mechanism for superconductivity, at least in
Tl$_2$Ba$_2$CuO$_6$\cite{juergen,moler,Artem}. This immediately
raises the question of whether the incoherent $c$-axis transport
arises from 2-dimensional confinement of single charge carriers to
the copper oxide planes, and whether or not confinement persists
in the superconducting state.
\\
From an analysis of the infrared reflectivity spectra using the
2-fluid model, Tamasaku {\em et al.}\cite{tamasaku} have concluded
that for $T<T_c$ the $c$-axis quasi-particle scattering rate of
La$_{2-x}$Sr$_{x}$CuO$_4$ for $T<T_c$ decreases strongly with
temperature, and "the T dependence looks very similar to that for
the quasiparticle scattering rate in the $ab$ plane". This result
indicated that coherent quasi-particle transport is recovered
simultaneously with the occurrance of a finite critical current
along the c-direction. A different result was later obtained by
J.H. Kim {\em et al.}\cite{kim}. Experimentally the c-axis
transport was monitored via the temperature dependence of the real
part of the conductivity, $\sigma_c(\omega)$, which exhibited a
sharp drop below T$_c$. The drop in $\sigma_c(\omega)$ was
attributed "exclusively to the opening of a gap, {\em i.e.,
without} a change of the electronic scattering rate" of the
carriers along the $c$-direction\cite{kim}. Later, similar
behaviour of $\sigma_c(\omega)$ versus temperature has been
reported for Bi$_2$Sr$_2$CaCu$_2$O$_{8+\delta}$\cite{Maeda}, and
YBa$_2$Cu$_3$O$_{7-\delta}$\cite{Hosseini}.
For YBCO, which at
optimal doping is the least anisotropic in the normal state, the
rise in $\sigma_{ab}$ combined with the drop in $\sigma_{c}$ leads
to a conductivity anisotropy as large as 10$^4$ at low temperatures.
\\
In this paper we address the issue of `confinement'
experimentally. We study the degree of coherence of $c$-axis
quasi-particle transport in the superconducting state, the
temperature dependence of the c-axis plasma frequency, and the
role of intra-bilayer splitting by comparing the single-layer
and bi-layer materials within the same family of Tl-based
cuprates.
\\
Here we report new data on epitaxially grown Tl$_2$Ba$_2$CuO$_6$
and Tl$_2$Ba$_2$CaCu$_2$O$_8$ thin films, with  superconducting
transition temperatures of 80 and 98K respectively, established by
susceptibility measurements. Details of the sample preparation and
growth methods have been reported elswhere\cite{ren,Willemsen,eddy}.
Dimensions of the films in the $ab$ plane are typically 50-100
mm$^2$. We measured grazing incidence reflectivity in the far
infrared FIR region (17-700) cm$^{-1}$ using a Fourier transform
spectrometer. A grazing angle of 80$^{\circ}$ was chosen in order
to predominantly probe the $c$-axis response for $p$-polarization
of the light. Absolute reflectivities were obtained by calibrating
the reflectivities against the reflectivity of a gold film
deposited {\em in situ} on the sample without moving or rotating
the sample holder.
\\
In Fig.~\ref{graz} we display the grazing incidence reflectivity
of Tl2201 (Fig.~\ref{graz}b) and Tl2212 (Fig.~\ref{graz}a) in the
frequency region from 17-50 cm $^{-1}$ for various temperatures
below T$_c$. The spectra above 50 cm$^{-1}$ are dominated by the
$c$-axis phonons, and a detailed analysis is presented
elswhere\cite{Artem1}. The only absorption below 50 cm$^{-1}$
corresponds to the Josephson plasmon, a collective oscillation of
the Cooper-pairs along the $c$-axis. The observed resonant
frequencies ($\omega_{J}$) at 4K are 27.8 cm$^{-1}$ in Tl2201 and
25.6 cm$^{-1}$ in Tl2212, and they redshift as the temperature
approaches T$_c$ from below. In Fig.~\ref{magn} we present
p-polarized reflectivity for various temperatures below
T$_{\text{c}}$, normalized to the 110K spectrum (which is
essentially featureless). A magnetic field of 0.35T was sufficient
to shift the resonance out of our spectral window. The strong
redshift upon applying an external magnetic field confirms that
the absorption corresponds to the Josephson plasmon. Let us first
address the question of why $\omega_{J}$ is almost the same in
Tl2201 and Tl2212: in a model of superconducting planes weakly
coupled by  Josephson coupling, a separate resonance frequency is
associated with each link\cite{lt21}. As the only structural
difference between the two materials is the insertion of an extra
CuO$_{2}$ layer and a layer of Ba$^{2+}$ ions separating the two
CuO$_{2}$ layers, the same resonance frequency should appear for
the link across the thallium oxide blocking layer.
\\
We analyze the spectra using the Fresnel formula for uniaxial
crystals\cite{juergen}. The resonance is due to the zero-crossing
of $\varepsilon_{c}(\omega)$, for which we adopt the standard
two-fluid expression appropriate for the superconducting state
\begin{equation}
 \varepsilon_{c}(\omega)=\varepsilon_{S}(\omega)
  - \frac{c^2}{\lambda_c(T)^2\omega^{2}}
  + \frac{4\pi i \sigma_{qp}(\omega)}{\omega}
 \label{eq:difun1}
\end{equation}
where $\lambda_c$ is the c-axis penetration depth. Assuming the 
Josephson model to be applicable to this case, the pole-strength 
of the second (London) term, coming from the condensate, provides 
directly the critical current in the c-direction. 
Hence $1/\lambda_c(T)^2$ represents the superfluid
density appropriate for the c-direction. Oscillations of the
condensate are electromagnetically coupled to interband
transitions and lattice vibrations, which are represented by the
dielectric function $\varepsilon_{S}$. The third term arises from
currents due to thermally activated quasi-particles. The condition
for propagation of longitudinal modes in the medium along the
c-direction in the long wavelength limit, is that
$\varepsilon_{c}(\omega)=0$, which occurs at the Josephson plasma
frequency
$\omega_{J}=c/\{\lambda_{c}\sqrt{\mbox{Re}\varepsilon_{S}(\omega_{J})}\}$.
This is also the frequency of long-wavelenth plasma-polaritons
propagating along the planes, which one observes in optical
experiments.
\\
The $c$-axis optical conductivity at the screened plasma frequency
$\sigma_{qp}(\omega_J)$ determines the width of the plasma
resonance in our spectra\cite{juergen}. This follows from the fact
that the resonance lineshape is given by
\begin{equation}
  R_{p}(\omega) \approx 1 - \mbox{Im} e^{i\phi}
  \sqrt{ 1 -
  \frac{\sin^{2}\theta/\mbox{Re}\varepsilon_{S}}
  {\omega(\omega+i\Gamma)-\omega_J(T)^2} }
\end{equation}
where $\theta$ is the angle of incidence, and
$\phi=(\pi-\mbox{Arg}\varepsilon_{ab})/2$ is a weakly frequency
dependent phase-factor, ranging from $\phi = 0$ (ideal
superconducting response) to $\phi = \pi/4$ (metallic response),
and
$\Gamma=4\pi\mbox{Re}\sigma_{qp}(\omega_{J})/\mbox{Re}\varepsilon_{S}(\omega_{J})$
determines the resonance linewidth. In this expression the
imaginary part of $\sigma_{qp}$ has to be included in
$\mbox{Re}\varepsilon_{S}$, and vice versa. 
We see now that the line-width of the $c$-axis plasma resonance 
is directly proportional to $\mbox{Re}\sigma_{qp}$ at the resonance 
position. This does {\em not} imply a direct proportionality 
of the plasma resonance line-width to the quasi-particle scattering 
rate. In fact assuming here a Drude line-shape of the quasi-particle term, 
$\mbox{Re}\sigma_{qp}=ne^{2}m_{\perp}^{-1}\tau/(1+\omega^2\tau^2)$, 
the plasma resonance line-width
is proportional to $n_{qp}/\tau$ if $\omega_{J}\tau \gg 1$ and to
$n_{qp}\tau$ if $\omega_{J}\tau \ll 1$. In LSCO\cite{tamasaku,kim},
YBCO\cite{homes}, and recently Tl2201\cite{basov} where crystals
were available suitable for reflection spectroscopy on the
ac-crystal face, it can be seen directly from the conductivity
spectra that $\sigma_{c}(\omega)$ is essentially independent of
frequency in the superconducting state for frequencies below
$\simeq$ 100 cm$^{-1}$. This corresponds to the limit where
$\mbox{Re}\sigma_{qp}\propto n_{qp}\tau$. Hence the drop of the
plasma-resonance linewidth below the phase transition represents
the reduction of quasi-particle density at the Fermi energy due to
the opening of the superconducting gap\cite{tamasaku}.
\\
By fitting Eq.\ref{eq:difun1} to the measured spectra, using the
full Fresnel expression for the grazing reflectivity, we obtain
accurate values of the $c$-axis superfluid density,
$\lambda^2(0)/\lambda^2(T)$, and the optical conductivity at the
resonance frequency, $\sigma_{1}(\omega_{J})$, as a function of
temperature.
\\
In Fig.~\ref{sig} we present the temperature dependance of the
$c$-axis superfluid density, $\lambda^2(0)/\lambda^2(T)$ and the
$c$-axis superfluid density of LSCO of Ref.\cite{kim}. Fitting the
$c$-axis superfluid density to a powerlaw
$\lambda^2(0)/\lambda^2(T) = 1 - (T/T_0)^{\eta}$ for temperatures
below $0.5T_c$ we obtain an exponent $\eta = 2.1 $ for Tl2201,
$\eta = 2.4 $ for Tl2212. Hence our results show that there is
also no linear temperature dependence for the $c$-axis superfluid
density  in Tl2201, and Tl2212, as has been observed in several
other high T$_{\text{c}}$ materials.\cite{Hosseini,Xiang,Jacobs}
The behavior in both Tl-based compounds is close to quadratic.
\\
In Fig.~\ref{sig}b we present the $c$-axis optical conductivity
versus temperature for Tl2201, Tl2212 and LSCO. In all compounds
$\sigma_{c}$ exhibits a smooth drop below T$_{\text{c}}$. The drop
in $\sigma_{c}(T)$ has also been observed in YBCO and Bi2212, and
is very different from the temperature dependence of the
$ab$-plane conductivity, where $\sigma_{ab}(T)$ shows a large peak
below T$_{\text{c}}$, attributed to a rapid increase in the
quasiparticle lifetime. The absence of a subsequent rise of
 $\sigma_{c}$ at lower temperature indicates that effectively we
observe no enhancement of the quasiparticle life time for this
transport direction.
\\
Recently Xiang and Wheatley\cite{xiang} calculated the c-axis
response assuming d-wave pairing, and assuming a model for the
c-axis transport where momentum parallel to the planes,
$k_{\parallel}$, is conserved. Based on LDA band theoretical
results \cite{OKA}, they argued that for HTSC's with a simple
(non-body centered) tetragonal structure, $t_{\perp} \propto
(\cos(k_x a) - \cos(k_y a))^{2}$, leading to an exponent $\eta =
5$ at low temperature. Allowing some scrambling  of
$k_{\parallel}$ due to impurity scattering during the interplane
hopping, reduces this to $\eta = 2$ at very low temperatures,
crossing over to the $T^5$ above a temperature which depends on
the impurity scattering rate.
\\
Within the same line of thinking, Ioffe and Millis\cite{millis} recently
proposed that the  scattering lifetime has a strong dependence on
$k_{\parallel}$ along the two-dimensional Fermi-surface, namely
\begin{equation}
  \Gamma(k_{\parallel}) = \frac{\Gamma_{0}}{4}\sin^{2}(2\Theta) +
\frac{1}{\tau}
\end{equation}
where $\Theta$ is the angle of $k_{\parallel}$ relative to the
diagonal (`nodal') direction. For the in-plane response this leads
to an expression for the optical conductivity which fits the
in-plane experimental spectra rather well, and has the
prototypical property of the optimally doped cuprates that the
effective scattering rate
$1/\tau^{*}\equiv\omega\mbox{Re}\sigma/\mbox{Im}\sigma$ has a
linear frequency dependence if $\Gamma_{0}$ is assumed to be
temperature independent, while $1/\tau \propto T^{2}$.
\\
For the c-direction, we must take into account, that $t_{\perp}
\propto (\cos(k_x a) - \cos(k_y a))^{2}$, as has been done by
Xiang and Wheatley\cite{xiang}. The c-axis transport arises in
this picture from regions at the Fermi-surface well removed from 
the nodes of the gap. Due to the strong suppression of $t_{\perp}^{2}$ 
along the nodal directions, $\sigma_{c}(\omega)$ probes only regions 
away from the nodes, closer to the regions of maximum gap value. In
those regions of k-space $\hbar\Gamma(k_{\parallel}$ is large,
{\em } of order 1 eV. This places the c-axis optical conductivity
well inside the dirty limit\cite{marel}, and it  provides a simple
microscopic argument why an analysis of the c-axis optical
conductivity in the superconducting state, using the
Mattis-Bardeen expressions for $s$-wave superconductors compares
so well to the experimental data\cite{kim}.
\\
The question of how to understand the large difference in scattering
properties and gap observed along two different optical axes, has
thus been shifted to a strong dependence of the scattering rate on
the position along the Fermi-surface (which is a {\em
phenomenological} assumption) and a strong dependence of the
hopping parameter on $k_{\parallel}$ (which is a straightforward
result of band-theory). The `confinement' in the sense of an
anomalous renormalization of the transport properties due to
many-body effects, is in this picture associated with the
anti-nodal regions in $k$-space. Near the nodal-regions the
confinement is dominated by a straight-forward single-particle
effect on $t_{\perp}(k_{\parallel})$, which follows directly from
band-theory. However, the anti-correlation between these two kinds
of confinement seems hardly a coincidence, and certainly deserves
further attention. This leaves open the question of the
microscopic implications of this phenomenological model.
The correspondence between $\Delta(k_{\parallel})$ and $t_{\perp}$
has been previously attributed to the interlayer tunneling
mechanism\cite{chakravarty}. Likewise the correspondence between
$\Delta(k_{\parallel})$ and $\Gamma(k_{\parallel})$ may contain
important clues regarding the pairing mechanism leading to
superconductivity.
\\
In conclusion, we have  observed the $c$-axis Josephson plasma
resonance in  Tl2201 and Tl2201. The values of the resonant
frequencies in the two compounds are quite close, which indicates
that the weakest link is the inter-cell link. In a magnetic field of
0.35 T the resonance is completely pushed out of our spectral
window, which is the expected behavior for the Josephson
collective exitation. From our data, we were able to extract the
c-axis superfluid density, and conductivity $\sigma_{c}(T)$. The
temperature dependence of both quantities indicate two-dimensional
confinement of the charge carriers.
%
\\$^{\ast}$Also P. N. Lebedev Physical Institute, Moscow.

\newpage
\newpage
\newpage
\end{document}